\documentclass[10pt,journal]{IEEEtran}

\usepackage{cite}
\usepackage{amsmath,amssymb,amsfonts}
\usepackage{graphicx}
\usepackage{xcolor}
\usepackage{xurl}
\usepackage[hidelinks]{hyperref}
\usepackage{siunitx}
\usepackage{booktabs}
\usepackage{multirow}
\usepackage{tabularx}
\usepackage{threeparttable}
\usepackage[shortlabels,inline]{enumitem}
\usepackage{tikz}

\usetikzlibrary{positioning, arrows.meta, shapes}
\tikzset{>={Latex[width=2mm,length=2mm]}}

\setenumerate[1]{leftmargin=1.0cm}

\begin{document}

\title{Benchmarking Anomaly Detection Across Heterogeneous Cloud Telemetry Datasets}

\author{
    Mohammad~Saiful~Islam 
    and~Andriy~Miranskyy%
    \thanks{Mohammad Saiful Islam and Andriy Miranskyy are with the Department of Computer Science, Toronto Metropolitan University, Toronto, ON, M5B 2K3, Canada. Email: \{mohammad.s.islam, avm\}@torontomu.ca.}%
}

\markboth{}%
{Islam and Miranskyy: Benchmarking Anomaly Detection Across Heterogeneous Cloud Telemetry Datasets}

\maketitle

\begin{abstract}
Anomaly detection is important for keeping cloud systems reliable and stable. Deep learning has improved time-series anomaly detection, but most models are evaluated on one dataset at a time. This raises questions about whether these models can handle different types of telemetry, especially in large-scale and high-dimensional environments.

In this study, we evaluate four deep learning models, GRU, TCN, Transformer, and TSMixer. We also include Isolation Forest as a classical baseline. The models are tested across four telemetry datasets: the Numenta Anomaly Benchmark, Microsoft Cloud Monitoring dataset, Exathlon dataset, and IBM Console dataset. These datasets differ in structure, dimensionality, and labelling strategy. They include univariate time series, synthetic multivariate workloads, and real-world production telemetry with over 100,000 features.

We use a unified training and evaluation pipeline across all datasets. The evaluation includes NAB-style metrics to capture early detection behaviour for datasets where anomalies persist over
contiguous time intervals.
This enables window-based scoring in settings where anomalies occur over contiguous time intervals, even when labels are recorded at the point level. The unified setup enables consistent analysis of model behaviour under shared scoring and calibration assumptions.

Our results demonstrate that anomaly detection performance in cloud systems is governed not only by model architecture, but critically by calibration stability and feature-space geometry. By releasing our preprocessing pipelines, benchmark configuration, and evaluation artifacts, we aim to support reproducible and deployment-aware evaluation of anomaly detection systems for cloud environments.

\end{abstract}

\section{Introduction} \label{sec:intro}
Cloud platforms are the backbone of modern services, supporting applications that operate at global scale~\cite{pourmajidi2018challenges,pourmajidi2019dogfoodinguseibmcloud,pourmajidi2021challenging}. Monitoring modern cloud systems is a complex and ongoing endeavour, driven by scale, heterogeneity, and continuously evolving workloads~\cite{pourmajidi2018challenges,pourmajidi2019dogfoodinguseibmcloud,pourmajidi2021challenging}. These systems generate large volumes of telemetry data across multiple layers, including API calls, service logs, infrastructure metrics, and traces~\cite{islam2021anomaly, islam2025anomaly}. This telemetry is used to monitor system health, detect operational issues, and ensure performance targets. When failures occur, cloud operators rely on early detection of anomalies to take corrective actions and prevent cascading failures~\cite{islam2020anomaly}.

Anomalies can take many forms, such as sudden spikes in latency, drops in request rates, or unusual patterns in resource usage. These are commonly categorized as point anomalies, contextual anomalies, or collective anomalies, each requiring different detection logic~\cite{islam2025anomaly}. In large-scale distributed environments, anomalies often propagate across system components and trigger incident tickets or user-visible service degradations. As cloud services continue to grow in size and complexity, anomaly detection systems must be robust, automated, and adaptive.

To meet this demand, a wide range of statistical and machine learning methods have been explored, including supervised, semi-supervised, and unsupervised approaches~\cite{chandola2009anomaly, ruff2018deep, goldstein2016comparative, zong2018deep, su2019robust, ren2019time}. These methods have been evaluated on datasets with varying levels of complexity, yet scalability and generalization remain core challenges. One critical barrier is the high dimensionality of telemetry data in large-scale cloud systems~\cite{aggarwal2001surprising}. As the number of signals increases, models face sparsity, noisy inputs, and irrelevant features that can obscure true anomalies~\cite{verleysen2005curse}. Traditional distance-based or clustering-based techniques often degrade under these conditions~\cite{bolton2002statistical}.

Recent work has explored deep learning models to address the challenges of anomaly detection in complex cloud systems. Sequential models such as GRU~\cite{cho2014learning} and TCN~\cite{bai2018empirical} have been used to capture temporal dependencies, while attention-based architectures such as Transformer~\cite{vaswani2017attention} model long-range interactions in multivariate telemetry. More recently, lightweight MLP-based models such as TSMixer~\cite{ekambaram2023tsmixer} have shown strong performance on high-dimensional time series. These models represent different architectural paradigms, including recurrent, convolutional, attention-based, and token-mixing approaches, and therefore introduce different inductive biases that may influence anomaly detection behaviour. Prior studies have reported promising results when these models are evaluated on individual datasets~\cite{islam2021anomaly, islam2025anomaly}.

Building on this foundation, we extend our focus to multi-cloud anomaly detection by benchmarking these models across four real-world telemetry datasets. Our goal is to assess their robustness, scalability, and sensitivity to structural differences in data, such as label sparsity and input dimensionality. By applying a unified pipeline and a consistent evaluation framework, this study aims to provide practical insight into how deep learning models behave across heterogeneous telemetry settings.

Some prior work has evaluated anomaly detection models across more than one dataset~\cite{audibert2020usad, tranad2022, xu2022anomaly}. However, these studies often focus on datasets drawn from similar domains or share common assumptions, such as univariate signals or controlled anomaly injections. As a result, there is limited understanding of how anomaly detection models behave across telemetry sources that differ substantially in structure, origin, and scale.

In operational cloud environments, anomaly detectors are rarely recalibrated continuously. Instead, detection thresholds are typically derived from historical data and deployed for extended periods during which workloads, traffic patterns, and system configurations evolve~\cite{islam2021anomaly,islam2025anomaly, hundman2018detecting}. As a result, the ability of a calibrated anomaly detection pipeline to generalize under distribution shift is a first-order operational concern. However, most existing evaluations implicitly tune calibration parameters on test data or assume stable telemetry distributions, masking failure modes that arise in real deployments.

In this paper, we aim to answer the following \textbf{research questions}:
\begin{itemize}
    \item \textbf{RQ1:} How do deep learning models behave when applied to telemetry datasets with different structures, dimensionalities, and noise characteristics?
    \item \textbf{RQ2:} How robust are these models under domain shift and scale differences across heterogeneous telemetry sources typical of evolving cloud services?
    \item \textbf{RQ3:} What trade-offs arise between detection accuracy, early detection behaviour, and generalization across datasets?
\end{itemize}

To answer these research questions, we evaluated anomaly detection models across heterogeneous cloud telemetry. The key \textbf{contributions} of this work are as follows:

\begin{enumerate}
    \item We present a large-scale cross-dataset evaluation of four representative deep learning architectures and one non-neural baseline for anomaly detection across heterogeneous cloud telemetry, spanning benchmark datasets and real-world production telemetry with over 100{,}000 features.
    
    \item We introduce a unified evaluation framework with strict training-only likelihood calibration, enabling deployment-realistic comparison of anomaly detection pipelines under no-leakage constraints.
    
    \item Through subgroup-level analysis and geometric case studies, we show that calibration generalization depends critically on feature-space stability rather than model architecture alone, and that static thresholds break down under distribution shift in high-dimensional cloud telemetry.
    
    \item Using a production cloud dataset, we demonstrate that near-zero or negative detection scores frequently arise from calibration instability rather than from failures in anomaly detection itself, highlighting a key risk in current benchmarking practices.
    
    \item We release (via \url{https://github.com/msi-ru-cs/anomaly-dataset-benchmark-public}) preprocessing pipelines, benchmark configurations, and evaluation artifacts to support reproducible and extensible evaluation of anomaly detection systems for cloud operations.
\end{enumerate}

Beyond benchmarking individual models, this work presents a practical evaluation framework for analyzing likelihood-based anomaly detection under strict training-only calibration in high-dimensional, multi-cloud telemetry.

The remainder of the paper is organized as follows. Section~\ref{sec:related} reviews the related literature. Section~\ref{sec:dataset} describes the telemetry datasets used in this study. Section~\ref{sec:method} presents the modelling approach and evaluation pipeline. Section~\ref{sec:results} reports the experimental results and analysis. Section~\ref{sec:discussion} discusses the implications of the findings and explicitly addresses the research questions. Section~\ref{sec:threats} discusses threats to validity. Finally, Section~\ref{sec:conclusion} concludes the paper and outlines future directions.

\begin{table*}[t]
\caption{Comparison of Telemetry Datasets Used for Benchmarking}
\label{tab:dataset_comparison}
\centering
\renewcommand{\arraystretch}{1.2}
\setlength{\tabcolsep}{8pt}
\begin{tabular}{p{3.2cm}p{2.8cm}p{1.6cm}p{1.8cm}p{4.6cm}}
\toprule
\textbf{Dataset} & \textbf{Source Domain} & \textbf{Dimensionality} & \textbf{Size (Rows)} & \textbf{Anomaly Labelling Method} \\
\midrule
Numenta Anomaly Benchmark (NAB)~\cite{lavin2015evaluating} &
AWS VM metrics, IoT sensors, office energy, NYC taxi demand &
1 &
1K–22K (per file) &
Time window-based labels for 58 univariate streams; labels tuned for early detection \\
Microsoft Cloud Monitoring Dataset~\cite{ren2021towards} &
Production cloud services (API latency, crash reports) &
1 &
$\sim$3K (per file) &
Binary anomaly labels based on service degradation events across 67 streams \\
Exathlon Dataset~\cite{jacob2021exathlon} &
Synthetic high-dimensional Spark workloads &
2.2K &
25K–30K (per trace) &
Six injected anomaly types (e.g., contention, slow input) in 93 traces; task-oriented labelling \\
IBM Console Dataset~\cite{islam2025anomaly} &
Real-world IBM Cloud frontend APIs (multi-cloud, multi-service) &
117K &
39,365 (5-min intervals) &
Ground truth from issue tracker, synthetic tests, and manual triage by IBM Ops team \\
\bottomrule
\end{tabular}
\end{table*}

\section{Related Work} \label{sec:related}
Anomaly detection in cloud systems presents unique challenges due to the scale, heterogeneity, and volatility of telemetry data generated by modern infrastructure~\cite{miranskyy2016operational, soldani2022survey, hagemann2020review}. Deep learning has shown promise in modelling temporal dependencies and uncovering complex patterns in telemetry data. However, most evaluations are performed on a single dataset, often limited in dimensionality or constrained to a specific application domain~\cite{soldani2022survey}. This narrow focus makes it difficult to assess how well anomaly detection models generalize across different telemetry sources and operational settings.

To better understand the limitations of current evaluation practices, we organize our review around four representative datasets: Numenta Anomaly Benchmark (NAB)~\cite{lavin2015evaluating, ahmad2017unsupervised}, the Microsoft Cloud Monitoring Dataset~\cite{MicrosoftCloudDataset:online, dang2021lookout}, Exathlon~\cite{jacob2021exathlon}, and the IBM Console Dataset~\cite{islam2024dataset, islam2025anomaly}. These datasets differ in dimensionality, anomaly labelling strategies, noise characteristics, and operational scope. Reviewing them side by side highlights persistent benchmarking gaps and motivates our decision to evaluate multiple deep learning models across these datasets using a consistent evaluation framework.

\subsection[aNumenta Anomaly Benchmark]{Numenta Anomaly Benchmark (NAB)\texorpdfstring{\footnotemark[1]}{}}
\footnotetext[1]{NAB dataset: \url{https://github.com/numenta/NAB} (Accessed July 2025).}
The Numenta Anomaly Benchmark (NAB) is one of the earliest and most widely adopted benchmarks for streaming time-series anomaly detection~\cite{lavin2015evaluating, ahmad2017unsupervised}. It introduced a window-based scoring function that emphasizes early anomaly detection and penalizes false positives, making it suitable for evaluating real-time monitoring systems.

Ahmad et al.~\cite{ahmad2017unsupervised} demonstrated the use of rolling likelihood-based scores with NAB and showed that online HTM-based detectors could achieve high sensitivity with low latency. TranAD~\cite{tranad2022} and USAD~\cite{audibert2020usad} further used NAB to benchmark deep models, including Transformer-based architectures and adversarial autoencoders, reporting improvements in early detection and robustness over classical baselines. While NAB remains influential as an evaluation benchmark, its univariate structure limits its applicability to multivariate telemetry scenarios. A detailed description of the dataset is provided in Section~\ref{sec:dataset}.

\subsection[Microsoft Cloud Monitoring Dataset]{Microsoft Cloud Monitoring Dataset\texorpdfstring{\footnotemark[2]}{}}
\footnotetext[2]{Microsoft dataset: \url{https://github.com/microsoft/cloud-monitoring-dataset} (Accessed July 2025).}
The Microsoft Cloud Monitoring Dataset captures telemetry streams collected from production cloud services and has been widely used to evaluate univariate anomaly detection methods in operational settings~\cite{ren2021towards}. It provides a realistic testbed for studying anomaly detection under service-level variability and incident-driven labelling.

Several recent studies have leveraged this dataset to assess detection accuracy and interpretability. Xu et al.~\cite{xu2022anomaly} proposed the Anomaly Transformer, demonstrating improved precision over recurrent baselines, while Dang et al.~\cite{dang2021lookout} applied LookOut to generate sparse and diverse explanations for detected anomalies. Despite its realism, the dataset’s univariate nature limits its suitability for evaluating scalable multivariate anomaly detection models. Dataset characteristics and preprocessing details are described in Section~\ref{sec:dataset}.

\subsection[Exathlon Dataset]{Exathlon Dataset\texorpdfstring{\footnotemark[3]}{}}
\footnotetext[3]{Exathlon dataset: \url{https://github.com/Exathlon/ExathlonBenchmark} (Accessed July 2025).}
The Exathlon dataset, introduced by Jacob et al.~\cite{jacob2021exathlon}, is a widely used benchmark for evaluating anomaly detection in large-scale distributed data processing systems. It was designed to study detection performance under controlled yet realistic workload perturbations and supports systematic comparison across modelling approaches.

Jacob et al.~\cite{jacob2021exathlon} evaluated traditional machine learning methods, autoencoder-based models, and variational autoencoders on this dataset, observing that deep learning models achieved high recall but faced challenges related to noise and interpretability. Subsequent work applied Transformer-based~\cite{zhou2022transformer_exathlon} and graph attention-based models~\cite{chen2023graph_exathlon}, reporting moderate improvements while highlighting the difficulty of handling high-dimensional, task-heterogeneous telemetry. A detailed description of the dataset structure and anomaly labelling is provided in Section~\ref{sec:dataset}.

\subsection[IBM Console Dataset]{IBM Console Dataset\texorpdfstring{\footnotemark[4]}{}}
\footnotetext[4]{IBM dataset: \url{https://doi.org/10.5281/zenodo.14062900}.}
The IBM Console Dataset is a publicly available, real-world telemetry dataset collected from the IBM Cloud Console platform~\cite{islam2025anomaly}. It was introduced to support the study of anomaly detection in large-scale production cloud environments and remains one of the few labelled datasets sourced from live cloud operations.

Islam et al.~\cite{islam2025anomaly} benchmarked GRU and ANN models on this dataset using likelihood-based scoring and NAB-inspired metrics. Their results showed that GRU consistently outperformed ANN, achieving higher recall and more stable training behaviour. However, performance was constrained by extreme dimensionality, noise, and label sparsity, exposing limitations of naive anomaly detection pipelines and motivating the need for scalable preprocessing, calibration, and robust evaluation strategies. A detailed description of the dataset structure and labelling process is provided in Section~\ref{sec:dataset}.

\subsection{Cross-Dataset Generalization and Benchmarking Gaps}
While each of the datasets described above has contributed to advances in time-series anomaly detection, most studies evaluate models in isolation. Prior work often optimizes model parameters separately for each dataset and reports metrics such as F1-score or NAB score without examining how well models generalize across different telemetry sources.

A small number of studies, including TranAD~\cite{tranad2022}, Anomaly Transformer~\cite{xu2022anomaly}, and MTAD-GAT~\cite{chen2021learning}, have evaluated models on more than one dataset. However, these evaluations are typically limited to univariate signals or synthetic multivariate traces and do not reflect the diversity of scale, structure, and noise observed in real-world telemetry.

To our knowledge, no prior work has systematically benchmarked deep learning models across datasets as diverse as NAB, Exathlon, Microsoft Cloud Monitoring, and the IBM Console dataset. This gap limits understanding of model robustness and sensitivity to changes in telemetry structure. Our work addresses this gap by evaluating four modern deep learning models across all four datasets using a consistent training and evaluation pipeline and by analyzing model behaviour under domain shift.

\section{Dataset} \label{sec:dataset}
In the related work, we discussed four telemetry datasets and their usage in prior anomaly detection studies. Here, we describe these datasets in technical detail, focusing on their scope, dimensionality, labelling, and data sources. This information helps characterize the heterogeneity of the datasets and motivates the need for cross-dataset evaluation. Unlike the related work, which focuses on prior usage, this section describes how the datasets are structured and used in our experimental setup.

\subsection{Numenta Anomaly Benchmark (NAB)}
The NAB dataset~\cite{lavin2015evaluating} contains 58 univariate time series from four distinct domains:

\begin{itemize}
    \item \textbf{Cloud Monitoring (AWS CloudWatch):} 16 time series capturing CPU, memory, and network metrics from virtual machines.
    \item \textbf{Energy Consumption:} 9 time series measuring office building HVAC and power usage.
    \item \textbf{Urban Mobility:} 10 time series recording public transportation data such as NYC taxi demand.
    \item \textbf{Sensor Streams:} 23 time series from industrial or lab environments (e.g., temperature, pressure).
\end{itemize}

Each file contains timestamped values and labelled anomaly windows suitable for early detection tasks. Series lengths vary from 1,000 to 22,000 observations. This benchmark has been widely adopted for evaluating real-time and univariate anomaly detection algorithms due to its domain diversity, ground truth rigor, and evaluation tooling. The full dataset is available at~\cite{nab_repo}.

NAB presents several practical challenges for anomaly detection.
While only a subset of the series are cloud-native, we retain the full dataset and report results at the subgroup level to preserve comparability with prior work and to assess robustness across heterogeneous telemetry domains.
Anomalies in NAB are often subtle, short-lived, and weakly separated from nominal behaviour, particularly in real-world series.
Prior benchmarking studies have shown that detection performance on NAB is highly sensitive to scoring methodology, anomaly window definitions, and threshold calibration, with conservative settings frequently suppressing detections for ambiguous cases~\cite{ahmad2017unsupervised,schmidl2022anomaly}.
In addition, window-based anomaly labelling introduces subjectivity in both evaluation and ranking, making results dependent on annotation and scoring choices rather than solely on detector behaviour~\cite{tatbul2018precision}.
As a result, zero detections or low scores can reflect dataset ambiguity rather than model failure under strict no-leakage evaluation.

\subsection{Microsoft Cloud Monitoring Dataset}
The Microsoft Cloud Monitoring Dataset~\cite{ren2021towards} includes 67 univariate telemetry streams collected from a range of production-grade online services. These metrics capture time-series traces of API latency, system health scores, CPU usage, crash frequency, and service availability. Each time series is labelled with binary anomaly indicators derived from real-world incident reports, degradation signals, and error logs~\cite{dang2021lookout}. Sampling frequencies vary across series (1-minute to 1-hour), reflecting operational heterogeneity across services. 

The dataset poses challenges such as varying series lengths, inconsistent label density, and intermittent missing values. It offers a valuable benchmark for evaluating point anomaly detection methods in realistic settings. We normalize each stream independently and use standard sliding windows for training and evaluation.

\subsection{Exathlon Dataset}
The Exathlon Dataset~\cite{jacob2021exathlon} simulates high-dimensional cloud workloads by repeatedly executing Apache Spark jobs under controlled conditions. It consists of 93 long-running telemetry traces (each ~25K rows), capturing ~2,200 signals per timestamp. These include low-level system metrics (CPU, disk I/O, memory), Spark-level states (stages, executors, tasks), and JVM statistics (garbage collection, heap size).

Anomalies are synthetically injected following a well-defined taxonomy that spans six types: resource contention, misconfiguration, failure injection, scheduling delay, slowness, and abnormal termination. Each injected anomaly is timestamped, facilitating fine-grained ground truth evaluation. The dataset supports trace-level testing of multivariate anomaly detectors and serves as a rigorous, reproducible testbed for benchmarking under synthetic stress.

Exathlon also introduces practical challenges for benchmarking.
Traces are synthetically generated using injected faults under predefined workload scenarios, which yields clean labels but also introduces strong regularities that differ from real-world cloud telemetry.
In addition, the dataset is high-dimensional and large in scale, with individual application traces spanning tens of gigabytes, creating substantial memory and computational overhead during training and evaluation.
Prior large-scale benchmarking studies therefore evaluated only a small subset of Exathlon traces when assessing algorithmic scalability~\cite{schmidl2022anomaly}.
In contrast, we include all available Exathlon application traces in our evaluation, which required up to 96\,GB of memory to process on a dedicated local server.

Beyond computational cost, we observed systematic generalization effects during testing.
For certain applications, reconstruction-based models exhibit a geometric shift between training and test distributions, where test samples progressively diverge from the training centroid even outside labelled anomaly windows.
This shift results in elevated reconstruction error and strongly negative normalized NAB scores.
We analyze this behaviour in detail in Section~\ref{subsec:case_study_exathlon}, where geometric distance visualizations help explain performance degradation that is not attributable to detection failure alone.

\subsection{IBM Console Dataset}
The IBM Console Dataset~\cite{islam2025anomaly} represents a large-scale, organically labelled dataset collected from live IBM Cloud operations over a 4.5-month period. It comprises 39,365 samples recorded at 5-minute intervals across seven global data centers, yielding over 117,000 telemetry features per row. Metrics span request-response behaviour (e.g., latency, success rates), service endpoints, HTTP methods, and application-level components.

Features are structured into domain-driven templates (e.g., \texttt{5XX\_count}, \texttt{4XX\_avg}) that group metrics by semantic similarity. Anomalies are labelled using incident ticket logs, automated regression tests, and manual triage from the operations team. The dataset is inherently noisy, imbalanced, and high-dimensional, presenting challenges for both unsupervised and supervised detection. It enables evaluation of models under realistic multi-cloud service conditions and is publicly available on Zenodo~\cite{islam2024dataset}. To the best of our knowledge, this dataset is the only publicly available, high-dimensional, labelled telemetry dataset collected from a live production cloud environment, making it uniquely suitable for evaluating anomaly detection methods under realistic operational conditions.

Despite its realism, the IBM Console Dataset poses several practical challenges for benchmarking. The extreme dimensionality and large raw data volume make preprocessing and model training computationally expensive. In our experiments, even after domain-driven feature grouping, several subgroups remained highly dimensional, requiring substantial memory resources during training and evaluation. In addition, anomaly labels are sparse and provided as coarse time windows derived from incident logs, which makes likelihood calibration sensitive to the train--test split and to minor temporal misalignment. As also observed in our ICSE SEIP 2025 study, these characteristics can lead to unstable likelihood thresholds and conservative detection behaviour under strict training-only calibration, particularly in the presence of distribution shift between training and test periods.

\subsection{Dataset Summary}
To highlight the diversity across our selected datasets, Table~\ref{tab:dataset_comparison} provides a side-by-side comparison. It summarizes the domain origin, dimensionality, size, and labelling strategy of each dataset. This comparative view helps contextualize the challenges of generalization, scale, and labelling heterogeneity faced by anomaly detection models when evaluated across varied telemetry sources.

\section{Methodology} \label{sec:method}
This section describes the methodology used to benchmark deep learning models for anomaly detection across heterogeneous telemetry datasets. We adopt a unified experimental pipeline that separates model training from anomaly scoring and calibration. All models are trained using consistent architectural configurations and limited hyperparameter tuning, while anomaly detection performance is primarily governed by a shared likelihood-based scoring framework. This design allows us to evaluate how different model architectures behave under domain shift, scale variation, and labelling heterogeneity, while minimizing confounding effects from dataset-specific tuning.

\begin{figure*}[t]
\centering
\resizebox{0.95\textwidth}{!}{
\begin{tikzpicture}[
  font=\small,
  box/.style={draw, rectangle, rounded corners=2pt, minimum height=0.9cm, minimum width=3.2cm, align=center},
  arrow/.style={->, thick}
]

\node[box, fill=gray!10] (err) {Reconstruction Error};
\node[box, fill=blue!10, right=2.5cm of err] (lik) {Likelihood Score};
\node[box, fill=orange!12, right=2.2cm of lik] (thr) {Threshold Comparison};
\node[box, fill=red!8, right=2.2cm of thr] (det) {Anomaly Decision};

\node[box, fill=green!10, below=1.2cm of lik] (gt) {Ground-Truth\\ Anomaly Window};

\draw[arrow] (err) -- node[above]{Rolling windows} (lik);
\draw[arrow] (lik) -- (thr);
\draw[arrow] (thr) -- (det);
\draw[arrow, dashed] (gt) -- (det);

\end{tikzpicture}
}
\caption{Illustration of likelihood-based anomaly detection. Reconstruction error is converted into a likelihood score using rolling windows and compared against a calibrated threshold. Ground-truth anomaly windows are used only for evaluation.}
\label{fig:likelihood_example}
\end{figure*}
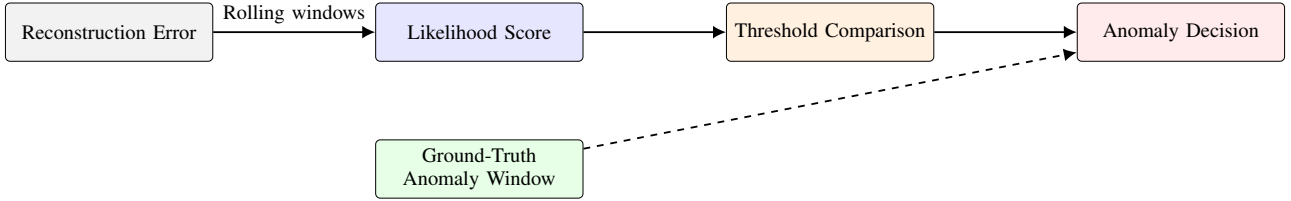

\subsection{Training, Validation, and Test Protocol}
\label{sec:method_split}

All experiments follow a fixed offline evaluation protocol.
For each telemetry time series, data are partitioned chronologically into a 70\% training split and a 30\% held-out test split.
From the training split, 10\% is reserved as a validation subset and used exclusively for early stopping and hyperparameter selection.

We adopt a strict no-leakage setup.
Model parameters are learned using the training split only.
Likelihood calibration parameters and detection thresholds are selected using the validation subset drawn from the training period.
No test labels or test statistics are used during model training or calibration.
After training and calibration, each model is executed once on the held-out test split to generate anomaly scores and detections.

This study focuses on offline benchmarking to ensure consistent and comparable evaluation across datasets and model architectures.
The use of disjoint training, validation, and test splits follows standard practice in machine learning and deep learning evaluation~\cite{bishop2006pattern,hastie2009elements,Goodfellow-et-al-2016}.
The experimental pipeline is configuration-driven and supports alternative split ratios.
Online and incremental training protocols can also be incorporated, as demonstrated in our prior work~\cite{islam2021anomaly}, although this study focuses on offline benchmarking for consistency.

\subsection{Modelling Framework}
We benchmark four representative deep learning architectures for time-series anomaly detection: GRU, TCN, Transformer, and TSMixer. These models were selected to cover a range of inductive biases, including recurrent, convolutional, attention-based, and token-mixing architectures. All models are implemented as reconstruction-based autoencoders using TensorFlow and are trained to minimize reconstruction error on normal telemetry patterns.

The \textbf{GRU} autoencoder is based on gated recurrent units~\cite{cho2014learning} and models temporal dependencies in sequential data. GRU-based models are effective in settings where temporal continuity is important and have been widely applied to anomaly detection in operational telemetry. The encoder maps fixed-length input windows into a latent representation, which is then reconstructed by a symmetric decoder.

The \textbf{TCN} autoencoder employs dilated causal convolutions to capture temporal dependencies over long horizons~\cite{bai2018empirical }. Compared to recurrent models, TCNs enable parallel computation and stable gradient propagation, making them suitable for modelling telemetry streams with abrupt changes and correlated signals across time.

The \textbf{Transformer} autoencoder relies on self-attention to model global temporal relationships within input windows~\cite{vaswani2017attention}. By attending to all positions in the sequence, the Transformer can capture long-range dependencies that may be difficult for recurrent or convolutional models to learn. However, its computational cost increases with sequence length and input dimensionality, which can affect scalability.

The \textbf{TSMixer} autoencoder is a lightweight MLP-based architecture that alternates between token mixing and channel mixing operations~\cite{ekambaram2023tsmixer}. Unlike attention-based models, TSMixer does not explicitly model temporal order but instead relies on mixing operations across time steps and features. This design enables efficient training on high-dimensional telemetry, particularly in settings with very wide input spaces.

In addition to these reconstruction-based deep models, we include \textbf{Isolation Forest} as a simple, non-neural baseline. Isolation Forest is an ensemble-based anomaly detector that identifies anomalies based on isolation depth induced by random feature sub-sampling and recursive partitioning~\cite{liu2008isolation,hariri2019extended}. Unlike autoencoder-based approaches, it does not learn a latent representation or rely on reconstruction error relative to a training centroid. Its inclusion provides a lightweight reference point for assessing whether observed performance differences arise from model complexity or from the calibration and scoring procedure itself.

Across all architectures, models are trained using fixed-length sliding windows and optimized using reconstruction loss. Architectural hyperparameters are kept consistent across datasets where possible, and model-specific tuning is intentionally limited. This ensures that observed performance differences are driven primarily by model inductive biases and dataset characteristics rather than aggressive hyperparameter optimization.

Likelihood calibration parameters are tuned using Bayesian hyperparameter optimization~\cite{snoek2012practical} implemented in the Weights \& Biases framework~\cite{wandb}. For each dataset subgroup and model, we run a Bayesian sweep of 100 optimization trials using only the training split. The search space is restricted to three likelihood parameters: the long window size, the short window size, and the detection threshold. No test data or test labels are used during tuning. Table~\ref{tab:likelihood_summary} summarizes the likelihood parameter values selected for the best-performing models across datasets.

\subsection{Anomaly Scoring and Likelihood Calibration}
All models produce pointwise reconstruction errors by comparing each input window with its reconstructed output. For multivariate inputs, reconstruction error is computed per feature and aggregated into a single error value per time step. This error sequence serves as the raw anomaly signal.

Following the NAB evaluation framework~\cite{lavin2015evaluating, ahmad2017unsupervised}, reconstruction errors are transformed into anomaly likelihood scores using a rolling-window normalization scheme. Two windows are maintained: a long window that captures the baseline error distribution and a short window that captures recent deviations. This formulation emphasizes early detection of sustained anomalies while reducing sensitivity to isolated noise spikes.

Formally, let $s_t$ denote the reconstruction error at time $t$. A long-term window of size $W$ is used to estimate the baseline error distribution, with empirical mean $\mu_t$ and standard deviation $\sigma_t$ defined as:
\begin{equation*}
\mu_t = \frac{1}{W}\sum_{i=0}^{W-1} s_{t-i},
\end{equation*}
\begin{equation*}
\sigma_t = \sqrt{\frac{1}{W-1}\sum_{i=0}^{W-1} (s_{t-i} - \mu_t)^2}.
\end{equation*}
A short-term window of size $W' \ll W$ is used to compute a recent mean $\tilde{\mu}_t$. The anomaly likelihood at time $t$ is then defined as:
\begin{equation*}
L_t = 1 - Q\left(\frac{\tilde{\mu}_t - \mu_t}{\sigma_t}\right),
\end{equation*}
where $Q(\cdot)$ denotes the Gaussian tail probability. An observation is classified as anomalous when $L_t$ exceeds a threshold.

Likelihood parameters, including the long window size, short window size, and detection threshold, are not fixed a priori. Instead, they are calibrated using automated hyperparameter sweeps. For each dataset subgroup, we perform a sweep over likelihood parameters and select the configuration that maximizes the NAB score on a validation slice drawn from the training period. This calibration procedure is applied uniformly across all models and datasets, ensuring that differences in detection performance arise from model behaviour rather than ad hoc threshold selection. Test labels are not used during likelihood calibration.

Anomalies are predicted when the likelihood score exceeds the calibrated threshold. For datasets with window-based annotations, only the earliest detection within each anomaly window is counted as a true positive, consistent with NAB scoring rules. False positives are penalized based on their temporal distance from annotated anomaly windows, and false negatives are counted once per missed window. 

By decoupling model training from likelihood calibration, the framework enables fair comparison across architectures and datasets. Models are trained to minimize reconstruction error, while detection sensitivity is governed by a shared likelihood-based scoring mechanism that adapts to dataset-specific noise and scale characteristics.

\begin{table*}[t]
\centering
\caption{Subgroup-level anomaly detection results on the test split (30\%) under strict training-only calibration. For each dataset subgroup, we report the likelihood calibration parameters that yield the best normalized NAB test score, together with the corresponding best-performing model.}

\label{tab:subgroup_results_with_params}
\begin{tabular}{llccccr}
\toprule
Dataset (\#Subgroups) & Subgroup & Long Win & Short Win & Threshold & Best Test NAB & Best Model \\
\midrule
\multirow{10}{*}{Exathlon (10)}
 & App1  & 108  & 24 & 0.9728 & 97.57 & Isolation Forest \\
 & App2  & 86  & 8  & 0.9985 & 73.02  & TSMixer \\
 & App3  & --- & --- & ---    & 0.00$^{\dagger}$ & ALL \\
 & App4  & --- & --- & ---    & 0.00$^{\dagger}$ & ALL \\
 & App5  & 109 & 7 & 0.9990 & 61.99  & TCN \\
 & App6  & 152 & 14 & 0.9984 & 25.26   & TCN \\
 & App7  & --- & --- & ---    & 0.00$^{\dagger}$ & ALL \\
 & App8  & 186 & 30  & 0.9933 & 0.00 & Isolation Forest \\
 & App9  & 421 & 19 & 0.9965 & 47.35 & Isolation Forest \\
 & App10 & 304  & 30  & 0.9959 & 49.03 & Isolation Forest \\
\midrule
\multirow{9}{*}{Microsoft (9)}
 & application-crash-rate-1 & 89 & 3 & 0.9406 & 31.76 & GRU \\
 & application-crash-rate-2 & 87 & 3 & 0.9984 & 35.70 & GRU \\
 & consumer-purchase-rate   & 162 & 3 & 0.9979 & 62.93 & TSMixer \\
 & data-ingress-rate        & --- & --- & ---    & 0.00 & ALL \\
 & ecommerce-api-incoming-rps & 463 & 3 & 0.9982 & 45.71 & Transformer \\
 & middle-tier-api-dependency-latency & --- & --- & --- & 0.00 & ALL \\
 & mongodb-application-rps  & --- & --- & ---    & 0.00$^{\dagger}$ & ALL \\
 & mongodb-machine-rps      & 99 & 3 & 0.9989 & 18.01 & GRU \\
 & service-unavailable      & --- & --- & ---    & 0.00$^{\dagger}$ & ALL \\
\midrule
\multirow{7}{*}{NAB (7)}
 & artificialNoAnomaly     & --- & --- & ---    & 0.00$^{\dagger}$ & ALL \\
 & artificialWithAnomaly   & 177 & 28 & 0.9291 & 11.06 & GRU \\
 & realAdExchange          & 188 & 3  & 0.9987 & 5.52  & Transformer \\
 & realAWSCloudwatch       & 475 & 9  & 0.9989 & 16.44 & TCN \\
 & realKnownCause          & 222 & 19 & 0.9992 & 2.30  & TSMixer \\
 & realTraffic             & 392 & 17 & 0.9710 & 20.26 & GRU \\
 & realTweets              & 74  & 10 & 0.9942 & 6.11  & Transformer \\
\midrule
\multirow{3}{*}{IBM (3)}
 & 5xx\_all\_metrics        & --- & --- & ---    & 0.00 & ALL \\
 & 5xx\_all\_metrics\_pca70 & 164 & 28 & 0.9972 & 0.00 & TCN \\
 & 5xx\_count               & --- & --- & ---    & 0.00 & ALL \\
\bottomrule
\end{tabular}

\vspace{0.5em}
\footnotesize{
$^{\dagger}$ A test NAB score of 0.00 corresponds to subgroups with zero ground-truth anomalies in the test split, where producing no detections is the correct model behaviour.
}
\end{table*}

\subsection{Data Preprocessing by Dataset}
Given the heterogeneity of telemetry sources, we apply dataset-specific preprocessing steps while maintaining a consistent windowing and evaluation strategy across all datasets. Preprocessing aligns raw telemetry with model input requirements while preserving the original anomaly semantics of each dataset.

\paragraph{NAB} Each univariate time series is processed independently. Values are normalized using statistics computed from the training split. Missing or malformed timestamps are removed during preprocessing to ensure temporally ordered input sequences. Anomaly labels are aligned using the benchmark-defined anomaly windows~\cite{lavin2015evaluating}. Fixed-length sliding windows are generated for model training and inference, and reconstruction error is computed at each time step.

\paragraph{Microsoft Cloud Monitoring} Each metric is treated as an independent univariate time series with binary anomaly labels~\cite{ren2021towards}. Series are normalized independently and segmented into overlapping windows. Missing or malformed timestamps are removed during preprocessing, and the remaining samples are temporally ordered to ensure contiguous input sequences. No value imputation or interpolation is applied, preserving the original data characteristics.

\paragraph{Exathlon} Each telemetry trace is processed independently as a multivariate time series~\cite{jacob2021exathlon}. During preprocessing, missing or malformed timestamps are removed, duplicate timestamps are resolved, and the remaining samples are temporally ordered to ensure contiguous input sequences. All available telemetry signals are retained without feature selection, preserving the original dimensionality of each trace. Anomaly labels are aligned using the injected anomaly intervals provided with the dataset.

\paragraph{IBM Console} Telemetry is collected at five-minute intervals and grouped using domain-driven feature templates~\cite{islam2025anomaly}. For this study, we focus on the \texttt{5XX\_count} feature group, which captures server-side failure behaviour and provides a controlled multivariate input setting. Features are normalized using statistics computed from the training split. Sliding windows are generated chronologically, and anomaly labels are aligned using incident windows derived from issue tracker data, automated test failures, and operational validation.

Across all datasets, preprocessing preserves temporal order and avoids label leakage between training, calibration, and evaluation splits. While preprocessing choices vary by dataset, the downstream modelling, likelihood scoring, and evaluation steps follow a unified pipeline to enable consistent cross-dataset comparison.

Figure~\ref{fig:likelihood_example} illustrates how reconstruction error is transformed into a likelihood score and compared against a calibrated threshold for anomaly detection.

\section{Results}
\label{sec:results}
This section presents anomaly detection results obtained under the strict no-leakage evaluation protocol described in Section~\ref{sec:method_split}. All models are evaluated using identical training, calibration, and testing splits, with likelihood thresholds calibrated exclusively on training data without access to test labels. Under these constraints, conservative likelihood calibration may legitimately yield zero or negative normalized NAB scores. Such outcomes generally indicate either the absence of reliable early detection or the high cost of false positives, rather than implementation errors or training instability.

\subsection{Analysis of the Results}
The results reported in this section reflect generalization behaviour under realistic deployment constraints imposed by the strict no-leakage evaluation protocol. Across datasets and models, conservative likelihood calibration leads to a wide range of normalized NAB scores, including zero or negative values for certain subgroups.

These outcomes highlight fundamental differences in how model architectures respond to dataset characteristics such as dimensionality, noise level, workload diversity, and labelling structure, rather than differences in training stability or implementation quality.

Table~\ref{tab:subgroup_results_with_params} summarizes normalized NAB scores across four telemetry datasets: NAB, Microsoft Cloud Monitoring, Exathlon, and the IBM Console dataset. For each dataset subgroup, we report the likelihood calibration parameters, the best test NAB score, and the corresponding best-performing model. Subgroups where all models achieve identical test scores, most commonly 0.00, are reported as complete ties. These ties typically reflect the absence of a detectable anomaly signal under the imposed calibration constraints, rather than equivalent detection capability across models.

Unless stated otherwise, all NAB scores reported in this paper refer to the normalized NAB score, where a null detector yields a score of 0 and an ideal detector yields a score of 100. Scores may become negative when false positives dominate. For readability, we use the term \emph{NAB score} to denote the normalized value throughout the paper.

\begin{figure}[t]
  \centering
  \includegraphics[width=\columnwidth]{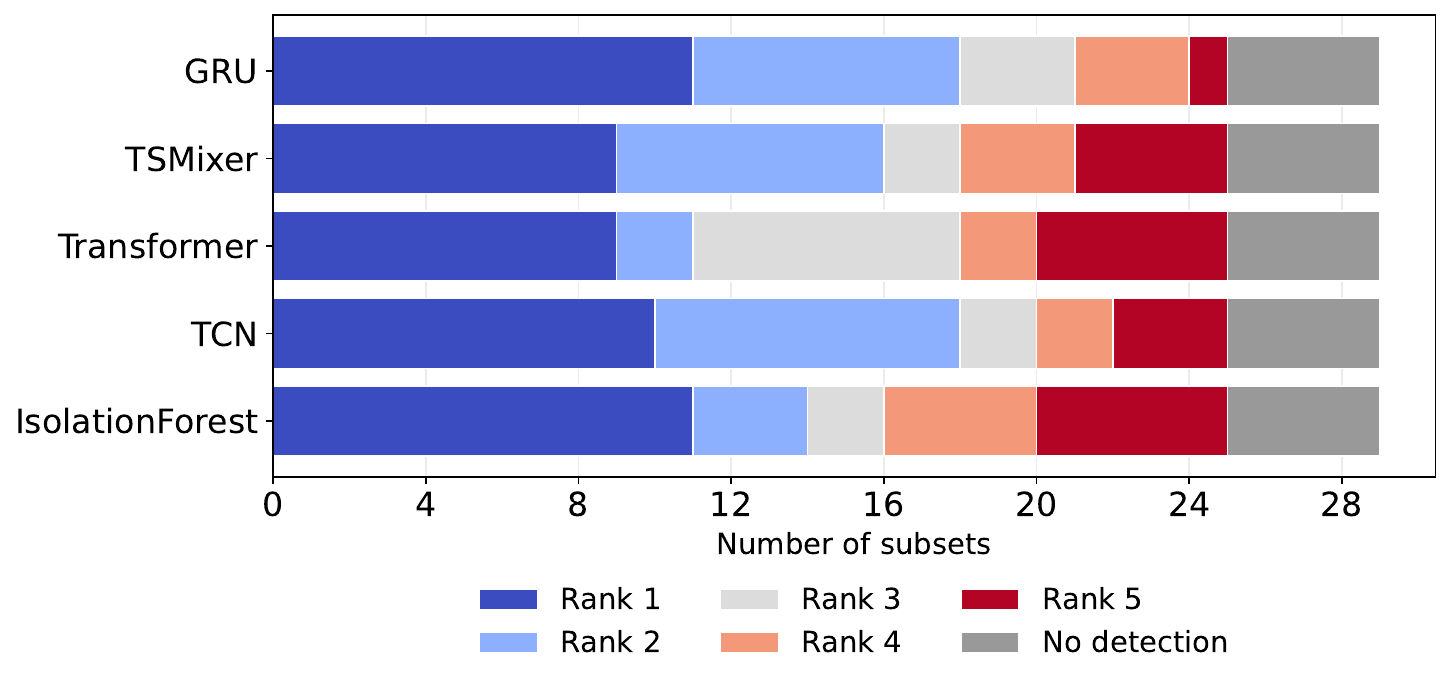}
  \caption{Ranking board over all 29 evaluated subsets using test-set NAB score. For each subset, models are ranked from best (Rank 1) to worst (Rank 5) based on their test-set NAB score using dense ranking. Subsets with no ground-truth anomalies, where all models correctly produce zero score, are counted as complete ties. Subsets with ground-truth anomalies in which no model produces any detection are excluded from Rank-1--Rank-5 counts and reported separately as a \emph{No-detection} category (grey).}
  \label{fig:rank_overall_ties_separate}
\end{figure}

Across all datasets, performance varies substantially across both models and subgroups. No single architecture consistently dominates all telemetry settings. This highlights the strong influence of dataset characteristics, including dimensionality, noise level, workload diversity, and labelling structure, on anomaly detection outcomes.

\subsection{Ranking Board Construction}
\label{subsec:ranking_board}

To summarize model robustness across heterogeneous telemetry subgroups, we transform per-subgroup test-set normalized NAB scores into a discrete ranking and then count how often each model attains each rank. For each subgroup, let $s_m$ denote the test-set normalized NAB score of model $m \in \{\text{GRU, TSMixer, Transformer, TCN, Isolation Forest}\}$, and let $GT$ denote the number of ground-truth anomaly windows in the test split. Higher scores are better.

We apply the following ranking rules per subgroup.

\paragraph{Case A (no anomalies in test)} If $GT = 0$ and all models produce $s_m = 0$, we treat this as correct non-detection and assign rank 1 to all models.

\paragraph{Case B (anomalies exist, but no model detects)} 
If $GT > 0$ and all models produce $s_m = 0$, we treat this as a \emph{no-detection} outcome. 
These cases are excluded from Rank-1 to Rank-5 counts and are reported as a separate 
\emph{no-detection} category in the ranking board plots.

\paragraph{General case} Otherwise, we rank models by descending $s_m$ using dense ranking, so tied models share the same rank and subsequent ranks are not skipped (e.g., $[10,10,3,-2] \rightarrow [1,1,2,3]$). While alternative schemes such as competition ranking are also common, we adopt dense ranking for reporting consistency and ease of aggregation across subgroups, following recent model benchmarking studies~\cite{bowles2024better,miranskyy2025feasibility}. This choice ensures that zero scores mixed with negative scores are ordered consistently (e.g., $[0,0,-5,-10] \rightarrow [1,1,2,3]$).

The resulting rank counts are aggregated across all subgroups (and optionally per dataset) to produce the ranking board visualizations reported in Section~\ref{sec:results}.

To summarize relative performance across heterogeneous subsets, we report a ranking-based view in Figure~\ref{fig:rank_overall_ties_separate}. For each subset, models are ranked by their test-set NAB score and aggregated by rank position. Cases where all models achieve identical scores, most commonly 0.00, are counted separately as ties (grey) to avoid imposing artificial ordering. A dataset-wise breakdown is provided in Appendix~\ref{app:ranking_by_dataset} (Figure~\ref{fig:ranking_board_ties_separate_by_dataset}) to localize where wins and ties originate.

On the NAB dataset, GRU and TSMixer achieve positive test scores across most series that contain ground-truth anomalies. These models exhibit stable early detection behaviour under window-based evaluation. Transformer-based models perform competitively on several series but show higher variability across subgroups. TCN achieves reasonable performance on selected series but is less stable overall compared to GRU and TSMixer under the same calibration strategy.

For the Microsoft Cloud Monitoring dataset, GRU demonstrates stable and consistently competitive test performance across multiple telemetry streams. TSMixer and Transformer perform well on selected subgroups but show increased sensitivity to likelihood calibration. Despite this variability, the relative ranking of models remains largely consistent across Microsoft subgroups, indicating robustness under moderate domain shift.

\begin{figure*}[t]
    \centering
    \includegraphics[width=0.8\textwidth]{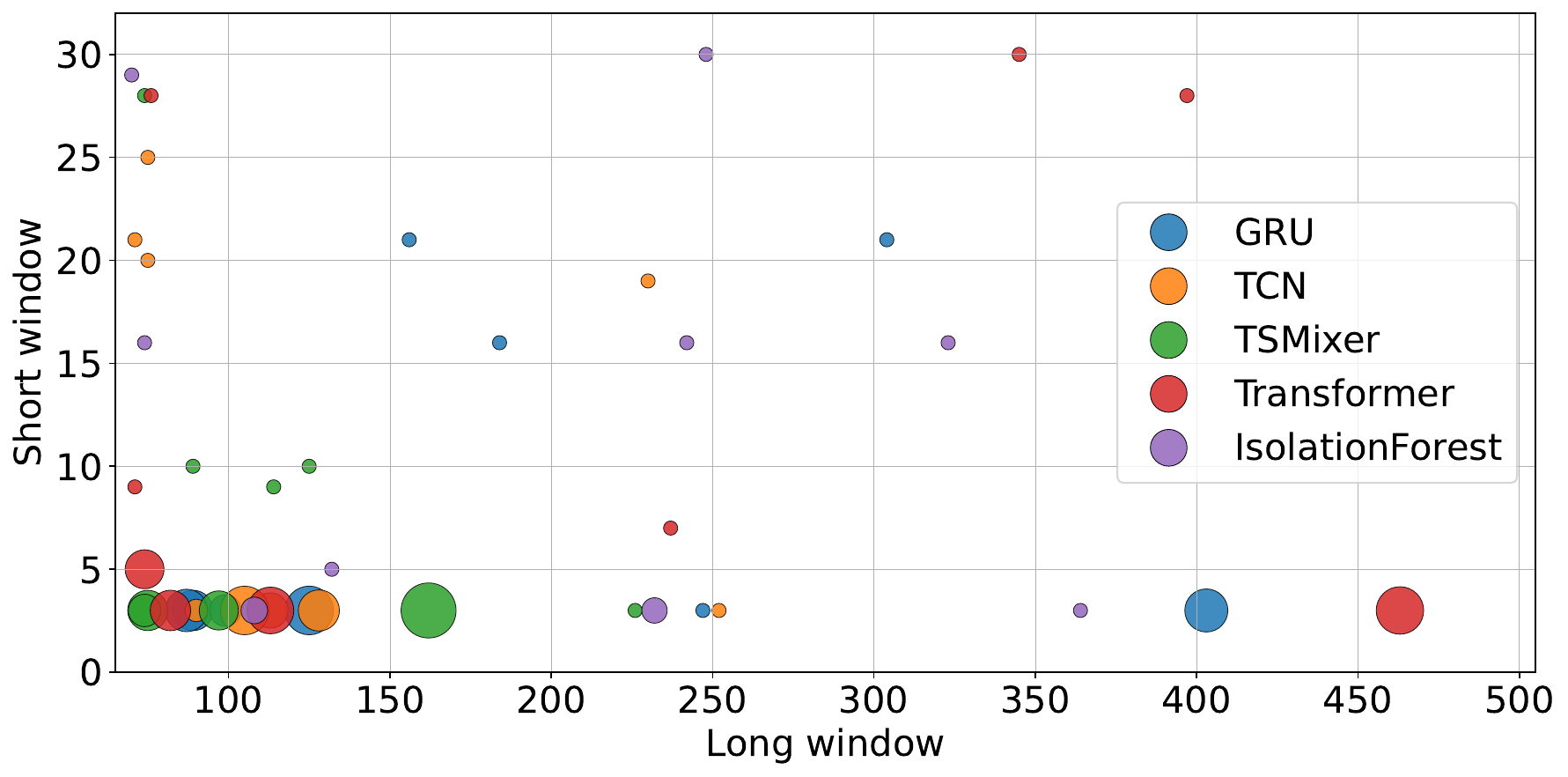}
     \caption{Calibration space for Microsoft telemetry.
The horizontal and vertical axes correspond to long and short likelihood windows.
Colors indicate models.
Bubble size is proportional to the normalized NAB test score, highlighting configurations that achieve stable positive test performance under conservative likelihood calibration.
}
    \label{fig:ms_calib_bubble}
\end{figure*}

To better understand this calibration sensitivity and stability, we visualize the likelihood calibration space for Microsoft telemetry in Figure~\ref{fig:ms_calib_bubble}, where this structure is most clearly observable.

\begin{table*}[!t]
\centering

\caption{Per-model normalized NAB scores on the test split (30\%) for all dataset subgroups. This table supports the ranking analysis (Figures~\ref{fig:rank_overall_ties_separate} and 
Appendix~\ref{app:ranking_by_dataset}--Figure~\ref{fig:ranking_board_ties_separate_by_dataset}) and distinguishes cases where a zero score reflects either a complete tie (all models score 0.00) or a least-penalized outcome (e.g., IBM \textit{5xx\_all\_metrics\_pca70}, where TCN scores 0.00 while other models incur negative scores).}

\label{tab:per_model_test_scores}
\scriptsize
\begin{tabularx}{\textwidth}{ll*{5}{>{\raggedleft\arraybackslash}X}}
\toprule
Dataset & Subgroup & GRU & TSMixer & Transformer & TCN & Isolation Forest\\
\midrule
\multirow{10}{*}{Exathlon (10)}
 & App1  & -38.50  & -195.46 & -294.52 & -190.24 & 97.57 \\
 & App2  & 15.60   & 73.02   & 64.63   & 27.70 & 52.45  \\
 & App3$^{\dagger}$  & 0.00    & 0.00    & 0.00    & 0.00  & 0.00  \\
 & App4$^{\dagger}$  & 0.00    & 0.00    & 0.00    & 0.00  & 0.00  \\
 & App5  & 61.99   & 48.20   & 60.44   & 63.29  & 0.00 \\
 & App6  & 7.05    & 0.45    & -307.81 & 25.26  & -16.50 \\
 & App7$^{\dagger}$  & 0.00    & 0.00    & 0.00    & 0.00  & 0.00  \\
 & App8  & -127.33 & -407.51 & -196.95 & -89.72  & 0.00 \\
 & App9  & -7928.64& -3234.74& -6647.65& -8291.74  & 47.35 \\
 & App10 & -131.33 & -174.65 & -102.15 & 0.00  & 49.03  \\
\midrule

\multirow{9}{*}{Microsoft (9)}
 & application-crash-rate-1              & 31.76 & 31.68 & 29.34 & 23.13 & 11.25 \\
 & application-crash-rate-2              & 35.70 & 29.64 & 32.01 & 33.20 & 0.00 \\
 & consumer-purchase-rate                & 48.24 & 62.93 & 44.16 & 48.12 & 0.00 \\
 & data-ingress-rate                     & 0.00  & 0.00  & 0.00  & 0.00  & 0.00 \\
 & ecommerce-api-incoming-rps            & 36.83 & 19.07 & 45.71 & 6.95  & 9.93 \\
 & middle-tier-api-dependency-latency    & 0.00  & 0.00  & 0.00  & 0.00  & 0.00 \\
 & mongodb-application-rps$^{\dagger}$               & 0.00  & 0.00  & 0.00  & 0.00  & 0.00 \\
 & mongodb-machine-rps                   & 18.01 & 0.00  & 0.00  & 0.00  & 0.00 \\
 & service-unavailable$^{\dagger}$                   & 0.00  & 0.00  & 0.00  & 0.00  & 0.00 \\
\midrule

\multirow{7}{*}{NAB (7)}
 & artificialNoAnomaly$^{\dagger}$     & 0.00  & 0.00  & 0.00  & 0.00  & 0.00 \\
 & artificialWithAnomaly   & 11.06 & 0.00  & -2.98 & 0.00  & 0.00 \\
 & realAdExchange          & 0.00  & -9.62 & 5.52  & 2.77  & 0.00 \\
 & realAWSCloudwatch       & 5.82  & 9.83  & -13.18& 16.44 & -0.11 \\
 & realKnownCause          & 0.00  & 2.30  & -1.81 & 0.00  & 0.00  \\
 & realTraffic             & 20.26 & 18.27 & 3.85  & -1.00 & 0.00 \\
 & realTweets              & 0.00  & 0.00  & 6.11  & 0.00  & -4.03 \\
\midrule

\multirow{3}{*}{IBM (3)}
 & 5xx\_all\_metrics        & 0.00   & 0.00   & 0.00   & 0.00 & 0.00 \\
 & 5xx\_all\_metrics\_pca70  & -11.66 & -13.71 & -10.29 & 0.00 & 0.00 \\
 & 5xx\_count               & 0.00   & 0.00   & 0.00   & 0.00 & 0.00 \\

\bottomrule
\multicolumn{7}{p{0.95\linewidth}}{\footnotesize
$^{\dagger}$ A test NAB score of 0.00 corresponds to subgroups with zero ground-truth anomalies in the test split, where producing no detections is the correct model behaviour.
}\\
\end{tabularx}

\end{table*}

The Exathlon dataset represents a more challenging setting due to its high dimensionality and heterogeneous workload patterns. While all evaluated models achieve positive test scores on several traces, a subset of files exhibits strongly negative NAB scores for reconstruction-based models, as shown in Table~\ref{tab:per_model_test_scores}. These cases are primarily associated with sparse anomaly windows and extended normal operating periods, which amplify false-positive penalties under NAB scoring. As a result, Exathlon exhibits the highest score variance among all evaluated datasets.

On the IBM Console dataset, which captures ultra-high-dimensional production telemetry, none of the evaluated models consistently achieve positive test scores under strict training-only calibration. The PCA-reduced variant yields limited improvement but still results in near-zero performance. These outcomes reflect the combined effects of extreme dimensionality, severe label sparsity, and operational noise typical of production cloud telemetry.

Across all datasets, calibrating likelihood parameters on the training split results in lower absolute NAB scores compared to more permissive calibration strategies. However, the relative ordering of models remains largely consistent. This indicates that the observed performance differences are more closely associated with model architecture and dataset characteristics than with calibration choices at test time.

Taken together, the aggregate ranking behaviour highlights two architectures that consistently perform well across heterogeneous datasets, namely GRU and TCN. Recurrent architectures such as gated recurrent units (GRU) networks have been widely adopted for anomaly detection in cloud and operational telemetry, due to their ability to model temporal dependencies and evolving system dynamics~\cite{islam2021anomaly,islam2025anomaly,lavin2015evaluating}. In contrast, while temporal convolutional networks (TCNs) have been previously applied to time-series anomaly detection in generic, sensor-based, and industrial monitoring settings~\cite{bai2018empirical,munir2018deepant}, their use in cloud-scale operational telemetry remains limited. To the best of our knowledge, the effectiveness of TCN-based anomaly detectors has not been systematically studied under strict no-leakage evaluation protocols and likelihood-based scoring frameworks in cloud monitoring contexts.

\paragraph{Interpretation of zero NAB scores}
In Table~\ref{tab:subgroup_results_with_params}, zero-valued test scores are reported uniformly as 0.00, with additional notation used only to distinguish correct non-detection from failure cases. A test NAB score of zero can arise under different conditions and therefore requires careful interpretation.

First, when a dataset subgroup contains no ground-truth anomalies in the test split, producing no detections is the correct and expected behaviour. In such cases, all evaluated models yield a test NAB score of 0.00, and we report the best model as \textit{ALL}. These entries are marked with a dagger ($^{\dagger}$).

Second, when ground-truth anomalies are present but all evaluated models fail to detect them, a test NAB score of 0.00 reflects failure rather than correct behaviour. In these cases, we report the best model as \textit{ALL} to indicate a complete tie at 0.00, even though the outcome represents non-detection rather than successful performance.

Third, when all evaluated models incur negative NAB scores due to false-positive penalties, but one model achieves a test score of 0.00 by producing no detections, the zero score represents the least penalized outcome under NAB scoring. In such cases, the model achieving the zero score is reported as best-performing for completeness. This situation occurs in the IBM Console dataset for the \textit{5xx\_all\_metrics\_pca70} subgroup, where the TCN model avoids false detections while other models incur negative scores.

This distinction allows us to separate correct non-detection from failure cases and provides a more faithful interpretation of anomaly detection performance under strict training-only calibration.

Beyond subgroup-level performance, we analyze how likelihood calibration parameters vary across datasets and how sensitive test performance is to these choices. This analysis provides insight into whether a single calibration strategy can generalize across heterogeneous telemetry sources.

\begin{table}[!t]
\centering
\caption{Summary of likelihood calibration parameter ranges selected for best-performing models across dataset subgroups under strict training-only calibration. Ranges reflect variation across subgroups within each dataset.}
\label{tab:likelihood_summary}
\begin{tabular}{lrrr}
\toprule
Dataset & Long Window & Short Window & Threshold Range \\
\midrule
Exathlon  
 & 86--421 
 & 7--30 
 & 0.9728--0.9990 \\

Microsoft 
 & 87--463 
 & 3 
 & 0.9406--0.9989 \\

NAB       
 & 74--475 
 & 3--28 
 & 0.9291--0.9992 \\

IBM       
 & 164 
 & 28 
 & 0.9972 \\
\bottomrule
\end{tabular}
\end{table}

Table~\ref{tab:likelihood_summary} summarizes the ranges of likelihood calibration parameters selected for best-performing models across dataset subgroups. The results show substantial variation in optimal window sizes and thresholds across datasets, reinforcing the dependence of likelihood calibration on the dataset.

Exathlon exhibits the widest range of long-window values, reflecting the diversity of workload durations and anomaly patterns in the dataset. Microsoft telemetry favors shorter short-window sizes with moderate long windows, consistent with high-frequency operational metrics. NAB displays a broad spread across all three parameters, indicating heterogeneous temporal dynamics even within curated benchmark data.

In contrast, the IBM Console dataset consistently selects a single calibration configuration with relatively large window sizes and a high threshold. This behaviour reflects the extreme dimensionality and sparsity of anomalies in production telemetry, where stronger temporal smoothing and conservative thresholds are required to limit false positives.

\paragraph{Per-model test-set performance}
To expose per-model failure modes that are not visible in best-per-subgroup summaries, Table~\ref{tab:per_model_test_scores} reports normalized NAB scores on the test split (30\%) for every model and subgroup. This table highlights cases where Isolation Forest attains stable near-zero or positive scores while reconstruction-based neural models incur large negative penalties under NAB scoring, especially in Exathlon.

\begin{table}[!t]
\centering
\caption{Stability characteristics of test-set normalized NAB scores across datasets.
Statistics are computed over all per-model test scores reported in Table~\ref{tab:per_model_test_scores} (all models and all subgroups).
Negative scores indicate heavy penalties from false positives under NAB scoring.}
\label{tab:stability}
\begin{tabular}{lrrr}
\toprule
Dataset & Score Range & Std. Dev. & Negative Scores \\
\midrule
Exathlon  
 & -8291.74--97.57 
 & 1871.01 
 & 17 / 50 \\

Microsoft 
 & 0.00--62.93 
 & 18.28 
 & 0 / 45 \\

NAB        
 & -13.18--20.26 
 & 6.69 
 & 7 / 35 \\

IBM       
 & -13.71--00.00 
 & 4.96 
 & 3 / 15 \\
\bottomrule
\end{tabular}
\end{table}

Table~\ref{tab:stability} further characterizes the stability of test-set NAB scores across datasets. 
All statistics are computed over \emph{all per-model test scores} reported in Table~\ref{tab:per_model_test_scores}, spanning all models and subgroups. 
The NAB and Microsoft datasets exhibit relatively moderate score ranges and comparatively low variance, with few or no negative scores across models. 
In contrast, Exathlon shows extreme variability, including very large negative NAB scores for several reconstruction-based models on multiple subgroups. 
These negative values arise from heavy penalties for false positives under NAB scoring, particularly in traces with long normal operating periods and sparse anomaly windows.

The IBM Console dataset exhibits a different failure mode. While a small number of per-model test scores become mildly negative (Table~\ref{tab:stability}), the majority of outcomes cluster near zero, indicating conservative non-detection or least-penalized behaviour under NAB scoring. This reflects the extreme dimensionality and label sparsity of production telemetry rather than systematic false-positive accumulation.

Overall, these results suggest that likelihood calibration exhibits strongly dataset-specific behaviour under strict no-leakage evaluation. At the same time, training-only calibration yields stable and mostly positive test performance for many NAB and Microsoft subgroups that contain ground-truth anomalies, while exposing substantial instability on Exathlon and near-zero outcomes on IBM. Treating calibration as a separate, tunable component therefore provides a clearer and more faithful assessment of anomaly detection performance across heterogeneous telemetry datasets, while also revealing failure modes that would be obscured by aggregate reporting.

To better understand the observed variability, we next examine representative success and failure cases at the subgroup level. In particular, we analyze a high-dimensional Exathlon workload where likelihood calibration yields consistently positive test performance across models, as well as a contrasting Exathlon subgroup where models incur severe false-positive penalties under NAB scoring despite strong training-set behaviour. These case studies provide insight into when training-only likelihood calibration generalizes reliably and when it breaks down due to data characteristics such as sparsity, anomaly placement, and operating regime shifts. An additional analysis of failure modes on the IBM Console dataset is presented separately due to its extreme dimensionality and label sparsity.

\subsection{Case Study Analysis: Exathlon}
\label{subsec:case_study_exathlon}

To better understand the subgroup-level variability observed in Tables~\ref{tab:subgroup_results_with_params} and~\ref{tab:per_model_test_scores}, we conduct a focused case study on the Exathlon dataset. Exathlon is well suited for this analysis due to its high dimensionality, heterogeneous workloads, and large variation in test-set normalized NAB scores.

We select two representative Exathlon subgroups: \textit{app5} and \textit{app9}. These subgroups are chosen because they exhibit contrasting behaviour under the same anomaly detection setup. In particular, \textit{app5} consistently achieves high normalized NAB scores, while \textit{app9} yields strongly negative scores for reconstruction-based models. For both cases, we use the same GRU-based model, with likelihood parameters tuned on the training split and evaluated on the held-out test split. From each subgroup, we analyze one representative file that contains ground-truth anomaly windows and contributes significantly to the subgroup’s overall result.

\paragraph{Geometric representation of high-dimensional time series}
Each timestamp in Exathlon is represented by a high-dimensional feature vector with 2,283 metrics, making direct visualization impractical. Instead of analyzing individual dimensions, we characterize the data distribution using a scalar summary derived from the geometry of the original feature space \cite{aggarwal2015outlier,verleysen2005curse}.

Let $\mathbf{x}_t \in \mathbb{R}^d$ denote the feature vector at time $t$, where $d = 2283$, and let $\mathcal{T}_{\text{train}}$ denote the set of training timestamps with cardinality $N = |\mathcal{T}_{\text{train}}|$. We compute the training centroid as

\[
\boldsymbol{\mu}_{\text{train}} = \frac{1}{N} \sum_{t \in \mathcal{T}_{\text{train}}} \mathbf{x}_t .
\]
For each timestamp, we then compute the $L^2$ distance $D_t$ to the training centroid,
\[
D_t = \lVert \mathbf{x}_t - \boldsymbol{\mu}_{\text{train}} \rVert_2 .
\]
This produces a single scalar time series that quantifies how far each point lies from the region of feature space learned during training. Importantly, this distance is computed using the full high-dimensional representation, without dimensionality reduction or detector-specific scores. It therefore reflects distributional changes in the underlying data rather than detector-specific artifacts.

\begin{figure*}[t]
  \centering
  \includegraphics[width=\textwidth]{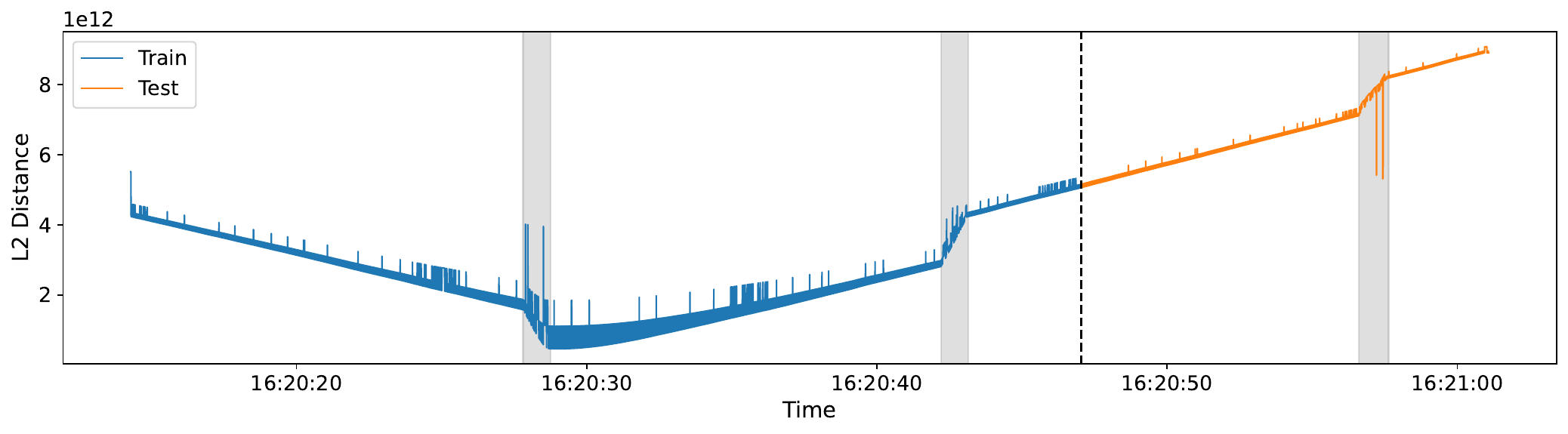}
  \caption{Distance to training centroid for \textit{app5\_\_5\_1\_100000\_64} using the GRU model. Each point represents the $L^2$ distance of a 2,283-dimensional feature vector to the training centroid. Training samples are shown in blue and test samples in orange. Shaded regions indicate ground-truth anomaly windows, and the dashed vertical line marks the train--test split. The strong overlap between training and test distances indicates stable geometry, enabling effective threshold generalization.}
  \label{fig:exathlon_app5_geometry}
\end{figure*}

\paragraph{Success case: \textit{app5} (stable geometry, high normalized NAB)}
For subgroup \textit{app5}, the representative file \textit{app5\_\_5\_1\_100000\_64} consistently achieves strong performance for the GRU model. When evaluated on the test segment, this individual file attains a normalized NAB score of 94.77, which is higher than the corresponding subgroup-level score reported in Table~\ref{tab:subgroup_results_with_params}.
As shown in Figure~\ref{fig:exathlon_app5_geometry}, the distance-to-training-centroid plot demonstrates strong geometric consistency between training and test segments. Distances remain on a comparable scale across the split, and anomaly windows appear as localized deviations from a stable baseline.

This geometric stability explains why a single likelihood threshold, tuned on the training data, generalizes effectively to the test data. In this case, the success of likelihood calibration reflects consistency in the underlying feature-space geometry rather than favorable threshold selection.

\paragraph{Failure case: \textit{app9} (unstable geometry, negative normalized NAB)}
In contrast, subgroup \textit{app9} exhibits severe degradation in performance. The representative file \textit{app9\_\_9\_4\_1000000\_78} produces a strongly negative normalized NAB score of $-158.57$ on the test segment, despite producing detections that overlap with labelled anomaly windows. These contrasts are visible in the per-model results in Table~\ref{tab:per_model_test_scores}, where \textit{app9} yields large negative scores for reconstruction-based models while Isolation Forest remains near-zero or positive.

\begin{figure*}[t]
  \centering
  \includegraphics[width=\textwidth]{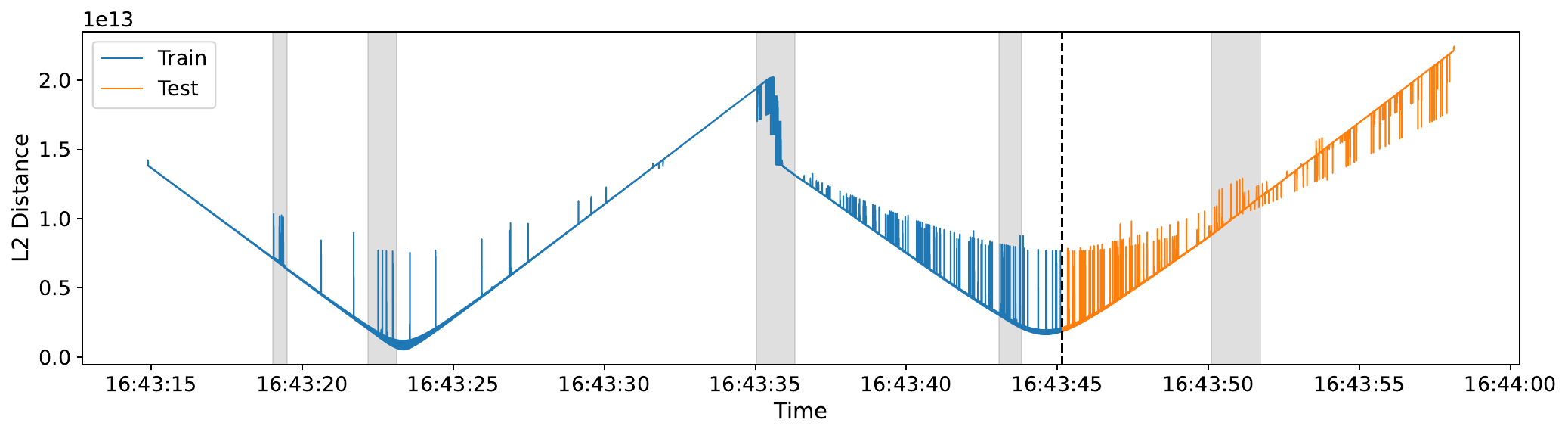}
  \caption{Distance to training centroid for \textit{app9\_\_9\_4\_1000000\_78} using the GRU model. While training samples remain close to the training centroid, test samples exhibit a systematic increase in distance, even outside anomaly windows. This geometric shift explains the large negative normalized NAB score despite correct anomaly detection.}
  \label{fig:exathlon_app9_geometry}
\end{figure*}

As illustrated in Figure~\ref{fig:exathlon_app9_geometry}, the distance-to-training-centroid plot reveals a pronounced geometric shift between training and test segments. While training samples remain close to the training centroid, test samples systematically drift farther away, even outside anomaly windows. As a result, normal behaviour in the test segment lies in a different region of the feature space than that learned during training. In this setting, anomaly windows are embedded within an already shifted baseline, causing a large number of false positives when using a fixed threshold and leading to catastrophic degradation in the normalized NAB score.

Notably, when evaluated under the same likelihood-based calibration protocol, Isolation Forest maintains near-zero or positive NAB scores for this subgroup by largely suppressing detections during the shifted test period, despite the presence of true anomalies.

\paragraph{Key insight}
Together, these two cases demonstrate that the generalization of likelihood-based thresholds in high-dimensional anomaly detection depends critically on the stability of feature-space geometry across train--test splits. When training and test data share a consistent geometric structure, as in \textit{app5}, likelihood calibration generalizes effectively. When the geometry shifts, as in \textit{app9}, no single static threshold can reliably separate normal and anomalous behaviour, regardless of detector architecture. This highlights a fundamental limitation of static thresholding under distribution shift in high-dimensional settings.

While this section focuses on Exathlon, we observe related failure modes in the IBM Console dataset due to its higher dimensionality and extreme label sparsity. A detailed geometric and subgrouping analysis for IBM telemetry is provided in Table~\ref{tab:per_model_test_scores} and Figure~\ref{fig:ibm_centroid_distance} in Appendix~\ref{app:ibm_case_study}.

Taken together, these results show that likelihood-based anomaly detection under strict training-only calibration exposes dataset-specific generalization limits that cannot be resolved through architecture choice alone, motivating a deeper discussion of representation, calibration, and evaluation design

\paragraph{Isolation Forest as a calibration-stability baseline}
To better understand whether the observed failure reflects limitations of anomaly detection or limitations of threshold calibration, we additionally evaluated Isolation Forest as a non-neural reference detector under the same likelihood-based scoring framework. Unlike reconstruction-based models, Isolation Forest does not rely on distances to a learned training centroid. Instead, it assigns anomaly scores based on random subspace partitioning and isolation depth~\cite{liu2008isolation, hariri2019extended}.

In high-dimensional settings such as Exathlon, this difference is consequential. Gradual global shifts between training and test distributions can inflate reconstruction error across thousands of features, even outside labelled anomaly windows. Such shifts do not necessarily produce correspondingly high isolation scores, as isolation depth is less sensitive to smooth, global displacement of the data distribution. As a result, Isolation Forest exhibits more conservative detection behaviour when thresholds are calibrated on the training split only, limiting false-positive accumulation under likelihood-based evaluation.

In the Exathlon subgroups where reconstruction-based neural models incur strongly negative normalized NAB scores, Isolation Forest therefore serves as a calibration-stability baseline, avoiding catastrophic score degradation rather than outperforming neural detectors. Detailed results are reported in Table~\ref{tab:per_model_test_scores}.

\section{Discussion}
\label{sec:discussion}

This section discusses the implications of the experimental findings reported in Section~\ref{sec:results}, with a focus on when likelihood-based calibration generalizes reliably and when it fails under strict training-only evaluation protocols.

\subsection{Addressing the Research Questions}

We now explicitly revisit the research questions posed in the introduction and answer them based on the empirical results presented in Section~\ref{sec:results}.

\textbf{RQ1: How do deep learning models behave when applied to telemetry datasets with different structures, dimensionalities, and noise characteristics?}
Our results show that model behaviour varies substantially across datasets and subgroups, and no single architecture consistently dominates across all telemetry settings. As shown in Table~\ref{tab:subgroup_results_with_params} and the ranking analyses in Figure~\ref{fig:rank_overall_ties_separate} and Figure~\ref{fig:ranking_board_ties_separate_by_dataset} (in Appendix~\ref{app:ranking_by_dataset}), different models emerge as best-performing depending on dataset characteristics such as dimensionality, anomaly density, and workload regularity. While GRU and TSMixer perform strongly on several NAB and Microsoft subgroups, performance diverges on high-dimensional Exathlon workloads and degrades uniformly on IBM Console telemetry. These findings indicate that anomaly detection performance is strongly dataset-dependent and cannot be inferred from architecture choice alone, as the same model may exhibit stable behaviour on one telemetry source and fail catastrophically on another under identical calibration constraints.

\textbf{RQ2: How robust are these models under domain shift and scale differences across heterogeneous telemetry sources typical of evolving cloud services?}
Robustness under domain shift is limited and emerges only when training and test data exhibit stable feature-space geometry, rather than as a function of model architecture or capacity. The Exathlon case study in Section~\ref{subsec:case_study_exathlon} demonstrates that when training and test data share consistent geometric structure, likelihood calibration generalizes effectively. In contrast, when test data exhibit systematic geometric drift relative to the training distribution, calibrated thresholds lead to excessive false positives and severe score degradation. Similar behaviour is observed on the IBM Console dataset, where extreme dimensionality and label sparsity yield conservative or near-zero outcomes across models. These results show that robustness under scale and domain shift depends primarily on distributional stability rather than detector complexity.

\textbf{RQ3: What trade-offs arise between detection accuracy, early detection behaviour, and generalization across datasets?}
Our evaluation reveals a trade-off between early detection sensitivity and calibration generalization under strict no-leakage constraints. 
Aggressive likelihood thresholds may improve early detection and apparent accuracy on stable datasets such as NAB, but they can amplify false positives under distribution shift, leading to negative NAB scores in Exathlon subgroups where test-time feature-space geometry diverges from the training distribution.
Conversely, conservative calibration suppresses false positives and improves generalization stability, but may result in non-detection or least-penalized outcomes on sparse or noisy telemetry. The comparison with Isolation Forest further illustrates this trade-off. Its conservative detection behaviour limits false positives under distribution shift and avoids catastrophic NAB score degradation, without being designed to optimize detection accuracy. Together, these findings highlight that early detection, accuracy, and generalization cannot be optimized simultaneously using static calibration in heterogeneous cloud telemetry.

\subsection{Generalization Behaviour of Likelihood Calibration}

Our results (Figures~\ref{fig:rank_overall_ties_separate} and Tables~\ref{tab:likelihood_summary}–\ref{tab:stability}) show that likelihood-based calibration is not inherently brittle, but that its generalization behaviour depends strongly on the underlying data distribution. Across multiple datasets, calibrated thresholds transfer effectively from training to test when the statistical and geometric properties of the data remain stable. In such cases, likelihood calibration acts as a reliable post-processing step that converts reconstruction errors into consistent anomaly decisions.

However, when training and test data exhibit distributional differences, likelihood calibration becomes highly sensitive. Even modest shifts in feature-space geometry can cause calibrated thresholds to produce excessive false positives, leading to severe degradation in evaluation scores. Importantly, this behaviour is observed consistently across model architectures, indicating that the limitation arises from data characteristics rather than model capacity.

Negative normalized NAB scores in our study (e.g., Exathlon subgroups in Table~\ref{tab:per_model_test_scores}) should be interpreted as indicators of calibration instability rather than detector incompetence. In particular, the Exathlon case study demonstrates that a model can correctly identify true anomaly windows while still producing a strongly negative NAB score due to excessive false positives in the test segment. As shown in the \textit{app9} subgroup, even when a true positive is detected, a systematic shift between training and test feature-space geometry causes normal test samples to be flagged repeatedly, leading to dominant false-positive penalties under NAB scoring. This behaviour highlights that negative scores primarily reflect calibration breakdown under distribution shift, rather than deficiencies in the underlying detection models.

\paragraph{Simple models versus complex architectures}
An important observation emerging from our results is that increased model complexity does not uniformly translate into improved anomaly detection performance. 
In aggregate, Isolation Forest attains a high number of first-place rankings, comparable to GRU (Figure~\ref{fig:rank_overall_ties_separate}). However, a dataset-level breakdown (Appendix~\ref{app:ranking_by_dataset}, Figure~\ref{fig:ranking_board_ties_separate_by_dataset}) reveals that this performance is highly uneven. Isolation Forest performs especially well on Exathlon, where it frequently achieves the top rank, but underperforms on NAB and Microsoft telemetry and ties with other models on the IBM Console dataset. This contrast highlights that simple, non-neural detectors can be effective in specific settings, but do not generalize uniformly across telemetry sources.

This finding relates to the analysis of Wu and Keogh~\cite{wu2021current}, who argue that many commonly used time-series anomaly detection benchmarks can create an illusion of progress by rewarding increasingly complex models without demonstrating robust generalization. Their study shows that, under careful evaluation, simple or even trivial baselines can match the performance of more sophisticated methods. In our results, comparable behaviour is observed in 1 out of 4 datasets, indicating that such effects may be dataset-specific.

Importantly, our goal is not to promote any specific model. Instead, these results highlight that anomaly detection performance is inherently dataset-dependent. Some telemetry sources benefit from expressive models with strong temporal inductive biases, while others are adequately handled by simpler detectors. This reinforces the absence of a one-size-fits-all solution and motivates dataset and application specific model selection supported by systematic benchmarking.

\subsection{Impact of High Dimensionality and Feature-Space Geometry}
High dimensionality plays a central role in amplifying calibration instability, as evidenced by the Exathlon results in Section~\ref{subsec:case_study_exathlon} and the near-zero outcomes observed for IBM Console subgroups in Table~\ref{tab:subgroup_results_with_params}.
 In datasets such as Exathlon and IBM Console, each timestamp is represented by thousands of metrics, increasing the likelihood that small distributional changes accumulate into large geometric shifts. As dimensionality grows, distances and likelihood estimates become increasingly sensitive to small deviations between training and test data, even when marginal feature statistics appear similar.

The geometric analysis presented in the Exathlon case study~\ref{subsec:case_study_exathlon} illustrates this effect clearly. By projecting high-dimensional feature vectors into a scalar distance-to-centroid representation, we show that test samples may systematically drift outside the region learned during training. In such settings, anomaly windows become embedded within an already shifted baseline, making it difficult for any fixed threshold to distinguish anomalous from normal behaviour. This observation provides a geometric explanation for why likelihood calibration may fail catastrophically despite correct anomaly detection at the window level.

\subsection{Implications for Benchmarking and Evaluation Protocols}

These findings have important implications for how anomaly detection systems are compared and evaluated. Many prior studies rely on evaluation protocols that implicitly tune calibration parameters on test data or otherwise mask instability under distribution shift. While such approaches may yield optimistic scores, they risk overlooking failure modes that are likely to occur in real deployments.

By enforcing strict training-only calibration, our evaluation protocol exposes these failure modes explicitly. Negative NAB scores in this context should therefore be viewed as informative signals of instability rather than as evidence of poor detector design. This perspective emphasizes the importance of subgroup-level analysis and discourages reliance on aggregate metrics alone when assessing anomaly detection performance.

\subsection{Dataset-Specific Challenges}

The severity and nature of calibration failures vary across datasets, as summarized in Tables~\ref{tab:likelihood_summary} and~\ref{tab:stability}.
In NAB and Microsoft Cloud datasets, moderate dimensionality and more stable operating regimes allow likelihood calibration to generalize in many cases. In Exathlon, mixed outcomes emerge, with some subgroups exhibiting stable geometry and others showing pronounced drift. The IBM Console dataset presents the most challenging scenario, combining extreme dimensionality, severe label sparsity, and noisy operational signals, which lead to conservative or least-penalized outcomes under strict evaluation.

\subsection{Practical Implications and Limitations}

From a practical perspective, our results suggest that static likelihood thresholds are risky in dynamic, high-dimensional environments. Operators deploying anomaly detection systems should expect calibration parameters to degrade over time and should monitor distributional changes explicitly. While our study does not propose adaptive calibration mechanisms, it highlights the need for geometry-aware monitoring and drift detection as complementary tools.

Finally, we note several limitations. Our analysis focuses on static, training-only calibration and does not explore adaptive or online thresholding strategies. In addition, while we provide a detailed geometric analysis for Exathlon, a comparable investigation for IBM Console telemetry is deferred to future work due to its additional complexities. Addressing these limitations represents a promising direction for further research.

\subsection{A Framework for Benchmarking High-Dimensional Multi-Cloud Telemetry}

Beyond individual model comparisons, this study contributes a practical framework for benchmarking anomaly detection systems on high-dimensional, multi-cloud telemetry. The framework combines strict training-only likelihood calibration, subgroup-level evaluation, and geometry-based diagnostic analysis to expose failure modes that are often hidden by aggregate metrics or permissive tuning strategies.

Rather than proposing a single detection model, our approach emphasizes how anomaly detection systems should be evaluated under realistic deployment constraints. In particular, it highlights the importance of separating calibration from detection, accounting for distributional stability across train--test splits, and interpreting zero or negative scores in context. This framework is applicable across heterogeneous cloud environments and provides a structured methodology for assessing robustness, generalization, and operational risk in large-scale telemetry settings.

\subsection{Implications for Cloud Operations}
From a cloud operations perspective, our findings suggest that static likelihood thresholds should be treated as fragile deployment artifacts rather than fixed properties of anomaly detection models. In moderately dimensional telemetry with stable operating regimes, training-only calibration can generalize effectively. However, in high-dimensional or evolving cloud environments, distributional drift can rapidly invalidate static thresholds, leading to excessive false positives or conservative non-detection.

These results motivate the need for operational safeguards such as geometry-aware drift monitoring, periodic recalibration policies, or adaptive thresholding mechanisms that explicitly account for feature-space evolution. While this study does not propose new calibration algorithms, it provides empirical evidence that calibration stability must be evaluated explicitly when deploying anomaly detection systems in large-scale cloud platforms.

\section{Threats to Validity}
\label{sec:threats}

We discuss potential threats to validity related to experimental design, dataset selection, evaluation methodology, and the generalizability of our findings.

\paragraph{Internal validity}
A primary threat to internal validity arises from the sensitivity of likelihood-based calibration to distributional differences between training and test data. To mitigate information leakage, all calibration parameters are tuned exclusively on training data. While this strict protocol may expose instability more clearly than approaches that tune on test data, it more closely reflects realistic deployment constraints. Differences in optimization convergence across models may also affect reconstruction error distributions; however, all models are trained using consistent procedures and evaluated under the same protocol.

\paragraph{Construct validity}
Our evaluation relies on the NAB scoring framework, which heavily penalizes false positives. While NAB is widely used and appropriate for streaming anomaly detection, its asymmetric penalty structure may amplify calibration failures under distribution shift. As a result, negative NAB scores should be interpreted as indicators of instability rather than an inability to detect anomalies. We mitigate this threat by complementing aggregate scores with subgroup-level and file-level analysis.

\paragraph{External validity}
The datasets used in this study span multiple domains, including benchmark datasets and real-world cloud telemetry. However, they may not capture all operational environments or workload characteristics encountered in practice. In particular, the IBM Console dataset exhibits extreme dimensionality and label sparsity, which may limit the generalizability of conclusions drawn from other datasets. Nevertheless, the diversity of datasets strengthens confidence that the observed trends are not entirely dataset-specific.

\paragraph{Conclusion validity}
Our conclusions are based on repeated experiments across multiple models, datasets, and subgroups. However, statistical significance testing is not performed, as the focus of this work is comparative benchmarking rather than formal hypothesis testing. To reduce the risk of overgeneralization, we emphasize qualitative patterns supported by consistent quantitative evidence, including stability metrics and case studies.

Overall, while these threats highlight limitations inherent to large-scale anomaly detection benchmarking, we believe our experimental design and analysis provide a robust and transparent assessment of likelihood-based calibration under realistic conditions.

\section{Conclusion}
\label{sec:conclusion}

In this paper, we presented a comprehensive benchmark of deep learning–based anomaly detection models across heterogeneous cloud telemetry datasets under a strict training-only calibration protocol. Our study focused on likelihood-based post-processing and evaluated its suitability across datasets with varying dimensionality, anomaly density, and degrees of distributional stability. By enforcing realistic evaluation constraints, we exposed both the strengths and limitations of likelihood calibration in operational settings.

Our results show that likelihood-based tuning can generalize effectively when training and test data exhibit stable feature-space geometry. However, in high-dimensional settings with distributional differences between training and test splits, static thresholds become fragile and may lead to severe false-positive penalties under NAB scoring. Through a detailed Exathlon case study, we demonstrated that negative NAB scores often reflect calibration instability rather than detector incompetence, even when true anomalies are correctly identified.

Taken together, these components form a practical evaluation framework for analyzing and benchmarking anomaly detection systems in high-dimensional, multi-cloud telemetry environments. These findings motivate a deployment-aware perspective on anomaly detection evaluation. Rather than viewing calibration as a secondary post-processing step, our results highlight it as a first-order systems concern that governs robustness, generalization, and operational risk in large-scale cloud telemetry. The evaluation framework presented in this study (combining strict training-only calibration, subgroup-level analysis, and geometric diagnostics) provides a practical basis for assessing anomaly detection pipelines in multi-cloud and high-dimensional environments.

From a practical standpoint, our study suggests that static likelihood thresholds should be applied with caution in dynamic cloud systems. Monitoring feature-space stability and subgroup-level behaviour is essential for determining when calibrated detectors remain reliable and when recalibration or additional safeguards are required. Near-zero or negative detection scores in such settings should be interpreted as signals of calibration breakdown rather than as evidence of detector failure.

Several directions for future work naturally follow from this study. In this work, we focus on evaluation under strict training-only calibration, rather than proposing adaptive detection or recalibration mechanisms. One promising extension is online or rolling-window calibration, where likelihood parameters are updated incrementally to adapt to evolving data distributions, albeit at the cost of increased training and operational complexity. Alternative calibration strategies, such as reconstruction-error normalization using adaptive mean and variance estimates or distance-based methods including Mahalanobis-style metrics, also merit further investigation. While we explored aspects of these approaches in preliminary experiments, a systematic evaluation of adaptive calibration under strict no-leakage constraints is left for future work. Together, these directions point toward geometry-aware and cost-sensitive calibration strategies for robust anomaly detection in large-scale cloud systems.

\bibliographystyle{IEEEtran}
\bibliography{references}

\clearpage
\appendix

\subsection{Dataset-wise Ranking Breakdown}
\label{app:ranking_by_dataset}

\begin{figure*}[t]
  \centering
  \includegraphics[width=0.95\textwidth]{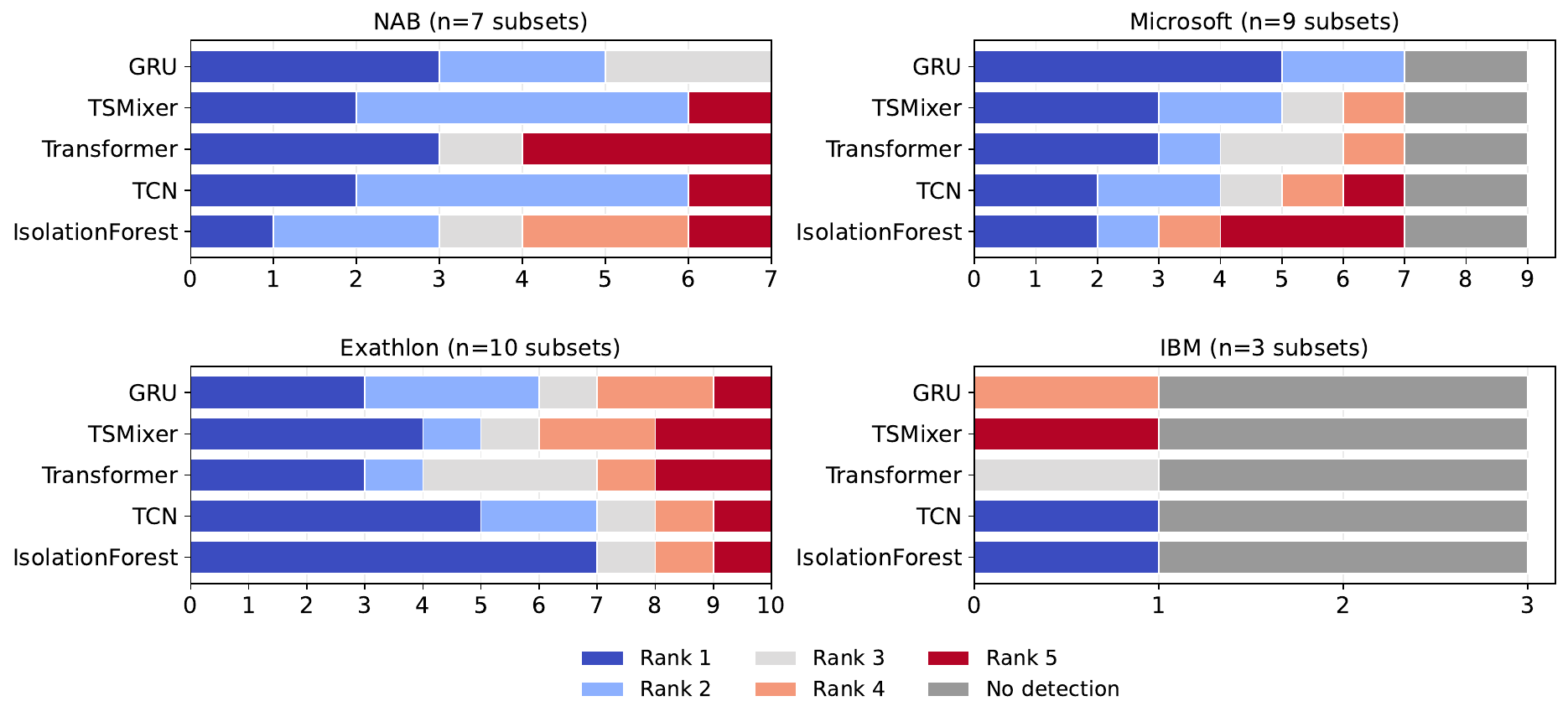}
  \caption{Dataset-wise ranking boards using test-set NAB score. Each panel shows the distribution of first through fourth place rankings for subgroups within a dataset family using dense ranking. Subgroups with no ground-truth anomalies, where all models correctly produce zero score, are counted as complete ties. Subgroups with ground-truth anomalies in which no model produces any detection are excluded from Rank-1--Rank-5 counts and reported separately as a \emph{No-detection} category (grey). This view is a drill-down across datasets of the overall ranking in Figure~\ref{fig:rank_overall_ties_separate}.}
  \label{fig:ranking_board_ties_separate_by_dataset}
\end{figure*}

Figure~\ref{fig:ranking_board_ties_separate_by_dataset} presents a dataset-wise breakdown of the ranking-based analysis introduced in the main results section. For each dataset family, models are ranked by their test-set NAB score on each evaluated subset, and the number of first through fourth place finishes is aggregated across subsets. Cases where all models achieve identical scores, most commonly zero, are counted separately as ties to avoid imposing artificial ordering.

This visualization helps localize where model wins, losses, and ties originate, and highlights differences in relative model behaviour across datasets with distinct telemetry characteristics. In contrast to the overall ranking view, this per-dataset perspective reveals dataset-specific patterns that may be obscured when aggregating results across heterogeneous sources.

\subsection{Case Study Analysis: IBM Console Telemetry}
\label{app:ibm_case_study}

\begin{figure*}[!t]
    \centering
    \includegraphics[width=\textwidth]{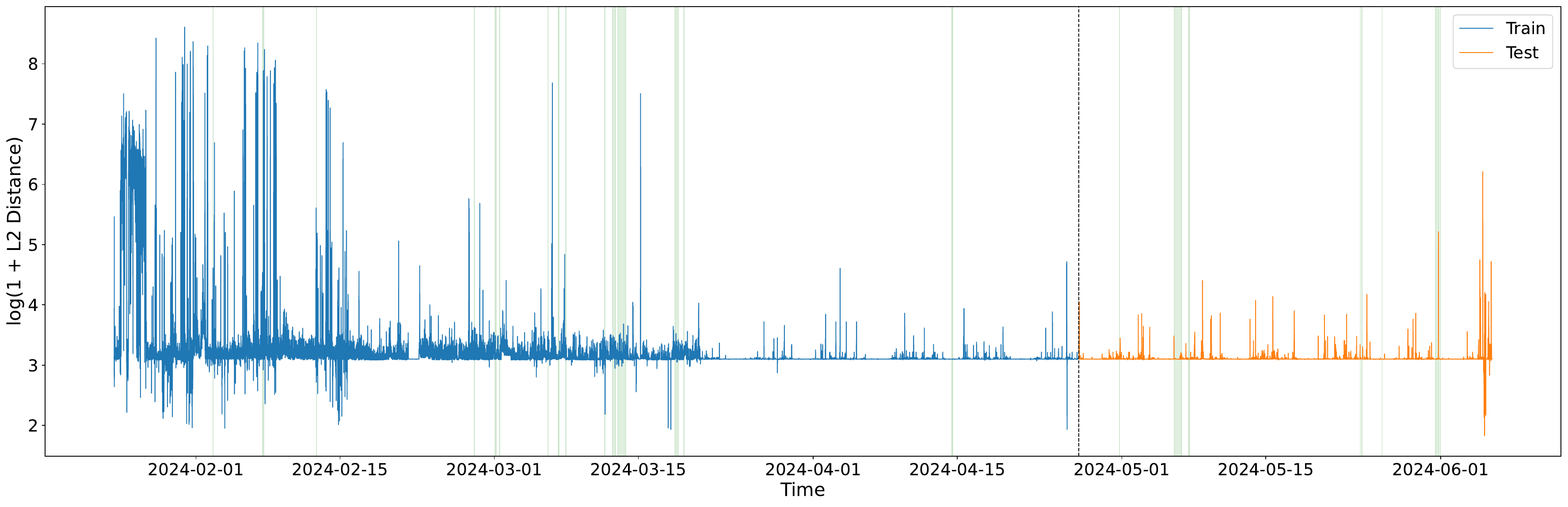}
    \caption{Log-scaled distance to the training centroid for IBM 5XX error telemetry using approximately 2,400 metrics. Shaded green regions denote labelled anomaly windows, and the dashed vertical line indicates the train--test boundary. Most anomaly windows remain close to the nominal operating region, while only a small number of isolated spikes exhibit large geometric deviations, leading to conservative likelihood thresholds under strict training-only calibration.}
    \label{fig:ibm_centroid_distance}
\end{figure*}

\paragraph{Geometric analysis of IBM telemetry}
Figure~\ref{fig:ibm_centroid_distance} plots the distance of each observation to the centroid of the training data, computed over approximately 2,400 5XX error-related metrics. Distances are shown on a $\log(1 + L^2)$ scale to improve readability by compressing rare extreme values. Shaded green regions indicate labelled anomaly windows, and the dashed vertical line marks the train--test boundary.

The visualization shows that most labelled anomalies remain close to the nominal operating region learned during training and do not correspond to large geometric deviations in the feature space. While a small number of isolated spikes exhibit substantial distance from the training centroid, these spikes are rare and do not consistently align with labelled anomaly intervals. This pattern holds across both training and test segments, indicating stable geometric behaviour under strict no-leakage evaluation.

As a result, likelihood thresholds calibrated exclusively on training data converge to conservative values that effectively suppress false positives but yield limited detections during evaluation. Under the NAB scoring framework, this behaviour leads to near-zero or least-penalized outcomes across models, reflecting the semantic and operational nature of production anomalies in IBM telemetry rather than a modelling or optimization failure.

We further observed similar behaviour under reduced representations, including PCA-based projections retaining 70\% variance and aggregated 5XX-count-only subgroups, indicating that the misalignment between geometric deviation and anomaly labels is not an artifact of feature dimensionality or aggregation.

\paragraph{Effect of structural subgrouping}
\label{app:ibm_structural_subgrouping}

To assess whether structural decomposition alone can mitigate the challenges posed by extreme dimensionality, we conducted auxiliary experiments using two additional subgrouping strategies on the IBM Console dataset: \emph{datacenter-wise} modelling and \emph{server--client-wise} modelling. These variants were implemented as controlled dimensionality reductions while preserving identical timestamps, anomaly labels, training splits, and evaluation protocol.

Despite this structured reduction, neither subgrouping strategy yielded consistently improved test-set performance under the strict 70--30 split. Across models, training behaviour remained stable, but test-time likelihood calibration continued to exhibit the same failure modes observed in the full-dimensional setting. These observations are consistent with the stability statistics reported in Table~\ref{tab:stability} and highlight a fundamental challenge in applying geometry-driven anomaly scoring to large-scale production telemetry.

\end{document}